\documentclass[lettersize,journal]{IEEEtran}
\usepackage{amsmath,amsfonts}
\usepackage{algorithmic}
\usepackage{array}
\usepackage{textcomp}
\usepackage{stfloats}
\usepackage{url}
\usepackage{verbatim}
\usepackage{graphicx}
\def\BibTeX{{\rm B\kern-.05em{\sc i\kern-.025em b}\kern-.08em		T\kern-.1667em\lower.7ex\hbox{E}\kern-.125emX}} 
\usepackage{balance} 

\usepackage{xargs}                  
 \usepackage{latexsym}
\usepackage[table,xcdraw]{xcolor}
\usepackage{tabularx}
\usepackage{booktabs}
\usepackage{amssymb, amsthm}
\usepackage{placeins}
\usepackage[font=footnotesize,labelfont=bf]{caption}
\usepackage{subcaption}
\usepackage{mathtools}
\usepackage{multirow}
\usepackage{cite, enumerate}
\usepackage[inline,shortlabels]{enumitem}
\usepackage{dsfont}
\usepackage{siunitx} 
\usepackage{stackengine,scalerel}
\usepackage{mathrsfs, eucal}
\usepackage{bm}
\usepackage{tikz}
\usepackage{circuitikz}
\usepackage{tikz-3dplot}
\usetikzlibrary{shapes, shapes.geometric, arrows, positioning}
\usepackage{tkz-euclide}
\tikzset{
    bs/.pic = {                                      
        \draw[line width = 1pt,-round cap] (0,0.4\R) -- (-0.2\R,-0.2\R);
        \draw[line width = 1pt,-round cap] (0,0.4\R) -- (0.2\R,-0.2\R);
        \draw[line width = 1pt] (-0.2\R,-0.2\R) -- (0.133\R,0);
        \draw[line width = 1pt] (0.2\R,-0.2\R) -- (-0.133\R,0);
        \draw[line width = 1pt,-round cap] (-0.133\R,0) -- (0.067\R,0.2\R);
        \draw[line width = 1pt,-round cap] (0.133\R,0) -- (-0.067\R,0.2\R);
        \draw[line width = 1pt,-round cap] (-0.213\R,0.4\R) -- (0.213\R,0.4\R);
        \draw[line width = 1pt,-round cap] \foreach \x in {-0.21, -0.14,...,0.22} {(\x\R,0.4\R) -- (\x\R,0.47\R)};
        \node (dim) at (0,0)  [align=center,minimum width=0.5\R,minimum height=1\R] {};
    },
}
\tikzset{
    user/.pic = {     
        \draw[rounded corners=0.02\R] (-0.09\R,-0.17\R) rectangle (0.09\R,0.17\R) ;                     
        \draw[fill=gray] (-0.08\R,-0.12\R) rectangle (0.08\R,0.12\R);                          
        \draw[rounded corners=0.005\R] (-0.04\R,0.14\R) rectangle (0.04\R,0.15\R);                     
        \draw (0,-0.14\R) circle (0.012\R) ; 
        \node (dim) at (0,0)  [align=center,minimum width=0.2\R,minimum height=0.3\R] {};
    }
}

\tikzset{
  drone/.pic = {
    \begin{scope}[scale=0.3]
      \fill[black] (-1.2,0) to[out=5,in=175] (1.2,0)
                     -- (0.7,-0.3) -- (-0.7,-0.3) -- cycle;
      \foreach \x in {-1.2,1.2} {
        \draw[very thick, black, line cap=round] (\x,0.1) -- (\x,0.4); 
        \draw[very thick, black, line cap=round] (\x-0.5,0.4) -- (\x+0.5,0.4); 
        \fill[black] (\x,0.4) circle (0.06); 
      }
      \fill[black] (-0.5,-0.3) -- (-0.9,-1) -- (-0.7,-1) -- (-0.3,-0.3) -- cycle;
      \fill[black] (0.5,-0.3) -- (0.9,-1) -- (0.7,-1) -- (0.3,-0.3) -- cycle;
      \fill[black] (-0.25,-0.75) rectangle (0.25,-1.25);
      \fill[red] (0,-1.0) circle (0.08);
    \end{scope}
  }
}

\tikzset{
    target/.pic = {
    \begin{scope}[scale=0.6]
        \tikzset{every path/.style={line width=1.3pt}}
        \draw
            (-1,0) -- (-0.75,0)
            .. controls (-0.65,0) and (-0.65,-0.1) .. (-0.6,-0.1)
            arc[start angle=270, end angle=450, radius=0.2cm]
            .. controls (-0.55,-0.1) and (-0.55,0) .. (-0.45,0)
            -- (0.45,0)
            .. controls (0.55,0) and (0.55,-0.1) .. (0.6,-0.1)
            arc[start angle=270, end angle=450, radius=0.2cm]
            .. controls (0.65,-0.1) and (0.65,0) .. (0.75,0)
            -- (1,0) -- (1,0.5) -- (-1,0.5) -- cycle;
        \draw[line width=1pt] (-0.6,0.5) -- (-0.4,0.9) -- (0.4,0.9) -- (0.6,0.5);
        \draw[line width=1pt] (-0.3,0.55) rectangle (0.0,0.85);
        \draw[line width=1pt] (0.1,0.55) rectangle (0.4,0.85);
        \fill (-0.6,-0.1) circle (0.02);
        \fill (0.6,-0.1) circle (0.02);
        \end{scope}
    }
}

\tikzset{
    repeater/.pic={
        \begin{scope}[scale=.8]
            \tikzset{every path/.style={line width=1.3pt}}
            \draw (-.5,0) rectangle (.5,.5);
            \draw (-.5,0) -- (-.5,.8);   
            \draw (.5,.5) -- (.5,.8);     

            \draw (-.7,1) -- (-.5,.8) -- (-0.3,1) -- cycle;
	     \draw (.3,1) -- (.5,.8) -- (0.7,1) -- cycle;
  	     \draw[->,thick, blue] (-.5,1) .. controls (-.3,2) and (.3,2) .. (.5,1) node [xshift= -4mm, yshift=3mm]{$\nu$};
        \end{scope}
    }
}

\tikzset{
    repeater_dual/.pic={
        \begin{scope}[scale=.78]
            \tikzset{every path/.style={line width=1.3pt}}
            \draw (-.6,-.2) rectangle (.6,.4);
            \draw (-.6,0) -- (-.6,.8);   
            \draw (.6,.4) -- (.6,.8);     

            \draw (-.9,1) -- (-.6,.8) -- (-0.3,1) -- cycle;
	     \draw (.3,1) -- (.6,.8) -- (0.9,1) -- cycle;
         \draw[->,thick, blue] (-.6,1) .. controls (-.4,1) and (-.4,1) .. (.25,1) node [xshift= -2mm, yshift=2mm]{$\alpha$};
	     \draw[<-,thick, red] (-.4,.8) .. controls (-.4,.8) and (-.4,.8) .. (.4,.8) node [xshift= -4mm, yshift=-2mm]{$\beta$};
        \end{scope}
    }
}

\tikzset{
    repeater_single/.pic={
        \begin{scope}[scale=.78]
            \tikzset{every path/.style={line width=1.3pt}}
            \draw (-.6,-.2) rectangle (.6,.4);
            \draw (0,.34) -- (0,.8);   

            \draw (-.3,1) -- (0,.8) -- (0.3,1) -- cycle;
         \draw[->,thick, blue] (-.4,1.2) .. controls (-.4,1.2) and (-.4,1.2) .. (.25,1.2) node [xshift= -3mm, yshift=2mm]{$\alpha$};
        \end{scope}
    }
}

\usepackage{accents}
\usepackage[normalem]{ulem}
\usepackage{tikz}
\usepackage{circuitikz}
\usetikzlibrary{shapes.geometric, arrows}
\DeclareMathAlphabet\mathbfcal{OMS}{cmsy}{b}{n}

\usepackage[acronym, shortcuts]{glossaries}
\makeglossaries
\usepackage[shortcuts]{glossaries}
\newacronym{6g}{6G}{sixth generation of mobile networks}
\newacronym{glrt}{GLRT}{generalized likelihood ratio test}
\newacronym{ris}{RIS}{reconfigurable intelligent surface}
\newacronym{se}{SE}{spectral-efficiency}
\newacronym{ncr}{NCR}{network-controlled repeater}
\newacronym{fdd}{FDD}{frequency-division duplexing}
\newacronym{mimo}{MIMO}{multiple-input multiple-output}
\newacronym{csi}{CSI}{channel state information}
\newacronym{bs}{BS}{base station}
\newacronym{ap}{AP}{access-point}
\newacronym{tdd}{TDD}{time-division duplex}
\newacronym{awgn}{AWGN}{additive white Gaussian noise}
\newacronym{ml}{ML}{maximum-likelihood}
\newacronym{psd}{PSD}{positive semi-definite}
\newacronym{iid}{i.i.d.}{independent and identically distributed}
\newacronym{rcs}{RCS}{radar cross-section}
\newacronym{evd}{EVD}{Eigen value decomposition}
\newacronym{svd}{SVD}{singular value decomposition}
\newacronym{ls}{LS}{least-square}
\newacronym{ils}{ILS}{iterative least-square}
\newacronym{cdf}{CDF}{cumulative distribution function}
\newacronym{uav}{UAV}{unmanned aerial vehicle}
\newacronym{wrt}{w.r.t}{with respect to}
\newacronym{aod}{AoD}{angle of departure}
\newacronym{aoa}{AoA}{angle of arrival}
\newacronym{music}{MUSIC}{multiple signal classification}
\newacronym{los}{LoS}{line-of-sight}
\newacronym{nlos}{NLoS}{non-line-of-sight}
\newacronym{mle}{MLE}{maximum-likelihood estimator}
\newacronym{pdf}{pdf}{probability density function}
\newacronym{isac}{ISAC}{integrated sensing and communication}
\newacronym{crlb}{CRLB}{Cramér-Rao lower bound}
\newacronym{ue}{UE}{user equipment}
\newacronym{lna}{LNA}{low-noise amplifier}
\newacronym{pa}{PA}{power amplifier}
\newacronym{sinr}{SINR}{signal-to-interference-plus-noise ratio}
\newacronym{snr}{SNR}{signal-to-noise ratio}
\newacronym{ra}{RA}{repeater-assisted}
\newacronym{ofdm}{OFDM}{orthogonal frequency division multiplexing} 

\usepackage[hidelinks, breaklinks, allcolors=blue]{hyperref}
\hypersetup{
	colorlinks   = true, 
	urlcolor     = blue, 
	linkcolor    = blue, 
	citecolor   = blue, 
}
\usepackage[capitalise,nameinlink]{cleveref}  
\usepackage{prettyref}
\crefformat{equation}{(#2#1#3)} 
\crefname{figure}{Fig.}{Figs.}
\crefname{section}{Sec.}{Secs.}
\crefname{algocf}{Algo.}{Algos.}
\Crefname{algocf}{Algorithm}{Algorithms}
\usepackage{float}
\usepackage[linesnumbered, ruled, vlined]{algorithm2e}
\SetCommentSty{mycommfont}

\SetCommentSty{xCommentSty}
\LinesNumbered
\SetSideCommentRight
\DontPrintSemicolon
\RestyleAlgo{algoruled}

\SetCommentSty{mycommfont}
\SetKwInput{KwInput}{Input}                
\SetKwInput{KwOutput}{Output}            
\SetKwInput{KwInt}{Initialization}            

\newenvironment{myfigure*}
{\begin{figure*}[!t]\centering}
	{\hrule\end{figure*}}

\newcommand{\ocirc}[1]{\ThisStyle{\ensurestackMath{%
  \stackon[.3pt]{\SavedStyle#1}{\SavedStyle\kern.05\LMpt\circ}}}}

\usepackage{fancyhdr}
\graphicspath{{./Figures/}} 

\usepackage{longtable}
\usepackage{array}
\usepackage{booktabs}
\usepackage{threeparttable}
\usepackage{tabularx}
\usepackage{xltabular} 
\usepackage{threeparttablex}
\definecolor{TableHeaderColor}{rgb}{0.74, 0.83, 0.9}

\allowdisplaybreaks
\begin{document}
\title{{Physical-Layer Aspects of Repeater-Assisted MIMO}}
\author{Diana P. M. Osorio,~\IEEEmembership{Senior Member,~IEEE}, Kohei Ueda,~\IEEEmembership{Graduate Student Member,~IEEE}, \\Anubhab Chowdhury,~\IEEEmembership{Member,~IEEE}, Hiroki~Iimori,~\IEEEmembership{Member,~IEEE}, Yuto~Hama,~\IEEEmembership{Member,~IEEE}, Koji~Ishibashi,~\IEEEmembership{Senior Member,~IEEE,} and  Erik G. Larsson,~\IEEEmembership{Fellow,~IEEE}
\thanks{Diana Osorio, Anubhab Chowdhury, and  Erik G. Larsson are with the Department of Electrical Engineering (ISY), Linköping University, 58183 Linköping, Sweden. Emails: diana.moya.osorio@liu.se, anuch87@liu.se, erik.g.larsson@liu.se}
\thanks{Kohei Ueda and Koji Ishibashi are with the Advanced Wireless \& Communication Research Center (AWCC), The University of Electro-Communications, Tokyo, Japan. Emails: kohei.ueda@ieee.org, koji@ieee.org}
\thanks{Hiroki Iimori and Yuto Hama are with Ericsson Research, Ericsson, Yokohama, Japan. Emails: yuto.hama@ericsson.com, hiroki.iimori@ericsson.com}
\thanks{This work was supported in part by the Swedish Research Environment ELLIIT, JST, CRONOS, Japan Grant Number JPMJCS24N1, and JST BOOST, Japan Grant Number JPMJBS2415.}
}
\maketitle
\thispagestyle{fancy}
\renewcommand{\headrulewidth}{0pt}
\fancyhf{}
\chead{\footnotesize\itshape This work has been submitted to the IEEE for possible publication. Copyright may be transferred without notice, after which this version may no longer be accessible.}

\begin{abstract} 
Network-controlled repeaters~(NCRs) are emerging as low-cost, band-selective active scatterers that can reshape the wireless propagation environment without backhaul or tight phase synchronization. We study physical-layer design for repeater-assisted multiple-input multiple-output~(RA-MIMO) networks, such as hardware impairments, gain and activation control, duplexing, channel-state information acquisition, and wideband delay effects. We then discuss integrated sensing and communication~(ISAC), where repeaters can extend non-line-of-sight~(NLoS) visibility, improve target observability, and enhance weak-echo detection. These gains, however, are limited by amplified noise, clutter, feedback, and calibration errors, motivating joint optimization and gain control frameworks for swarm repeater deployments within ISAC.
\end{abstract}

\section{Introduction}
\Gls{ncr} are frequency-selective, low-cost, low-power wireless devices that receive, amplify, and instantaneously retransmit wireless signals. Large plug-and-play deployments, termed \emph{swarms of repeaters}, act as \emph{active scatterers} that create amplified propagation paths while remaining transparent to the network~\cite{Jianan_TWC}. These repeaters extend coverage, provide macro-diversity, and improve channel rank in cellular massive \gls{mimo} systems~\cite{Sara_Vieira_Erik_Mag}. 

\begin{figure}[t]
    \centering
    \begin{subfigure}[t]{0.4\textwidth}
        \centering
        \includegraphics[width=.8\linewidth]{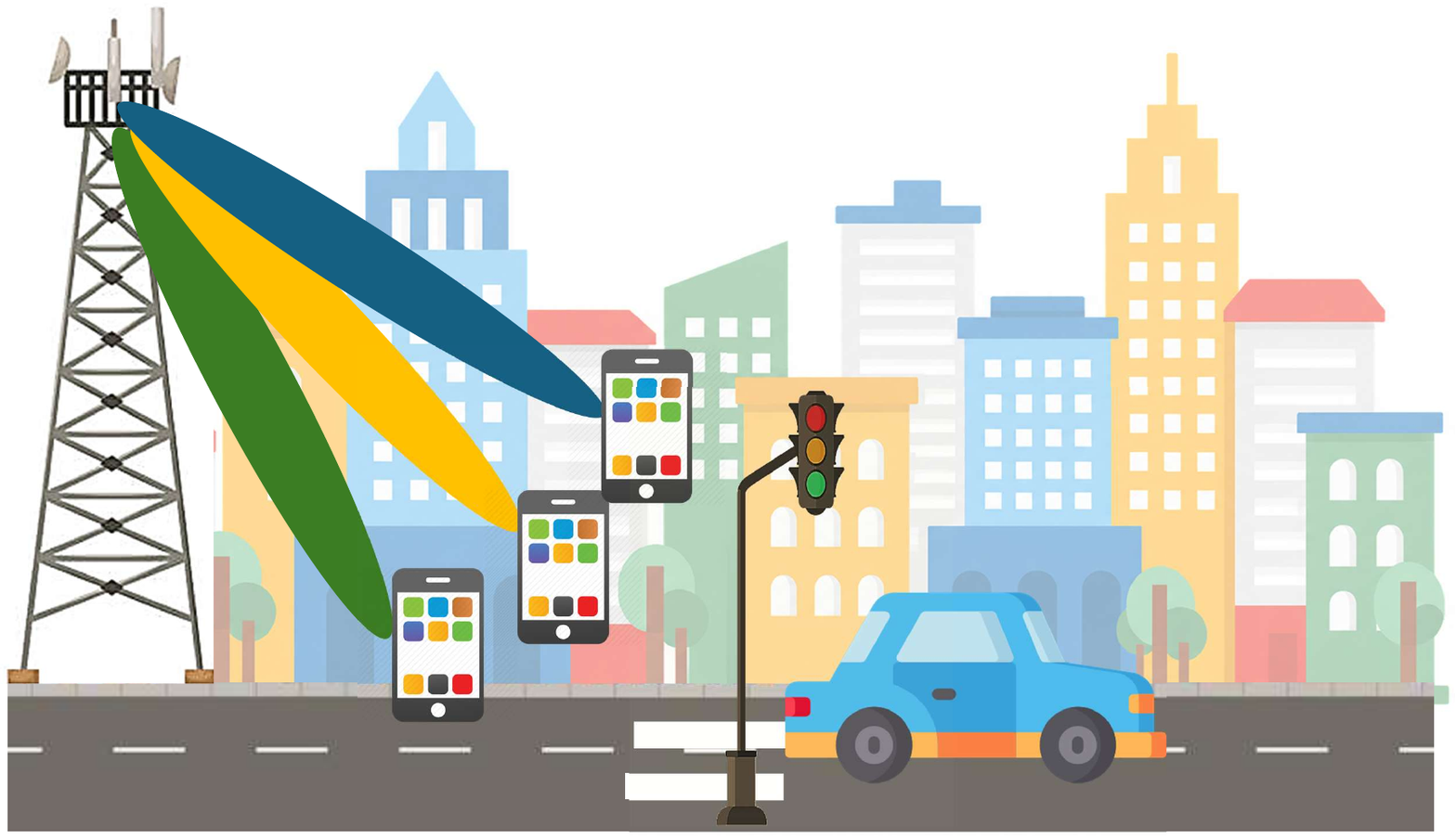}
        \caption{Massive MIMO network.}
        \label{fig:massive_mimo}
    \end{subfigure}
    \hfill
    \begin{subfigure}[t]{0.4\textwidth}
        \centering
        \includegraphics[width=.8\linewidth]{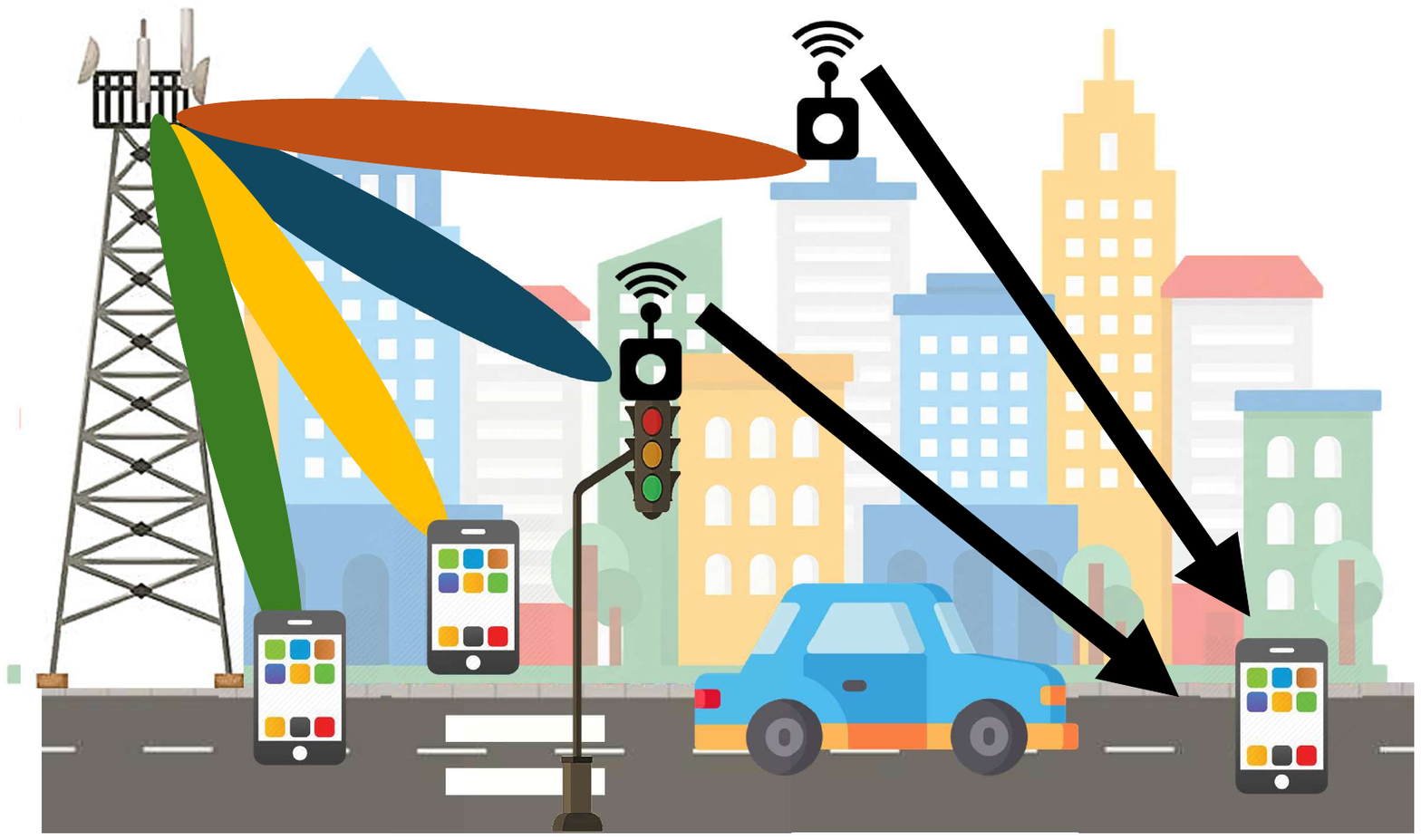}
        \caption{Repeater-assisted MIMO network.}
        \label{fig:ra_mimo_network}
    \end{subfigure}
    \caption{From Massive MIMO to RA-MIMO.}
    \label{fig:ra_mimo}
\end{figure}

Prior work showed that repeater swarms can achieve \gls{se} close to distributed \gls{mimo}~\cite{Sara_Vieira_Erik_Mag} without requiring backhaul or stringent phase synchronization. Repeater-assisted cellular systems also outperform \gls{ris}-assisted systems \cite{RIS_Makki_IEEE_Acess}, while offering sharp band selectivity. Essentially, unlike passive \gls{ris}, which reflect all operator bands equally and can cause uncontrollable inter-band interference, repeaters can isolate bands and are well-suited to multi-operator deployments.
In large swarms, repeaters may interact with one another. Stability criteria that prevent positive feedback within the swarm of repeaters have recently been derived in \cite{Jianan_TWC}. Within this stability region, repeater gains and transmit powers can be optimized for uniformly high throughput \cite{Jianan_TWC, Emil_repeater}, enabling large-scale deployment of repeaters; see \Cref{fig:ra_mimo} for illustration.

\begin{figure}[t]
    \centering    
    \includegraphics[width=.5\textwidth]{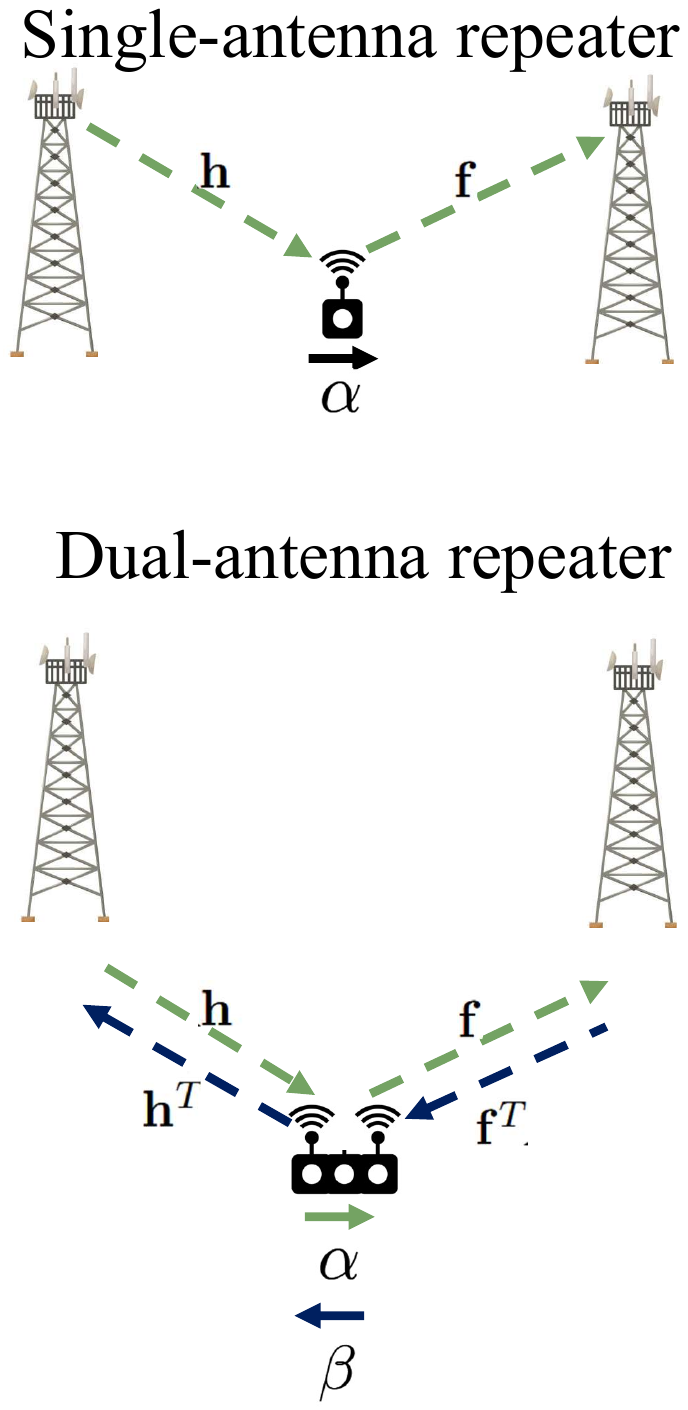}
    \caption{Different configurations for repeaters: single- and dual-antenna.}
    \label{fig:ra_mimo_config}
\end{figure}

\subsubsection{Propagation models for repeaters}\label{sec: prop_ch}
Repeaters can be single- or dual-antenna; see~\Cref{fig:ra_mimo_config}.
For single-antenna repeaters, the same antenna is used for simultaneous reception and transmission. The forward-link~(transmitter to receiver) channel via the repeater is the outer product of the channel from the transmit \gls{bs} to the repeater and the channel from the repeater to the receive \gls{bs}, multiplied by the forward amplification gain~(this is $\alpha$ in~\Cref{fig:ra_mimo_config}). Therefore, the repeater adds a rank-one component to the overall direct-link channel between the transmit and receive \glspl{bs}. 
Here, the reverse link over-the-air channel is reciprocal. 

For dual-antenna repeaters, the overall propagation channel also consists of the channel via the repeater and the direct-link channel, and the forward-link channel is similar to what we just described for a single-antenna repeater. However, in the reverse-link channel from the receive \gls{bs} to the transmit \gls{bs}, the reverse path amplification gain of the repeater~($\beta$ in~\Cref{fig:ra_mimo_config}) is multiplied by the outer product of the over-the-air channels. Therefore, for the repeater to be transparent,  dual-antenna repeaters need to be reciprocity calibrated~\cite{Erik_WCL}, which in turn enables \gls{tdd} reciprocity-based beamforming in the downlink.

\textcolor{black}{Our earlier magazine article~\cite{Sara_Vieira_Erik_Mag} established that repeater swarms can approach distributed-MIMO spectral efficiency. This article takes a different vantage point. Rather than arguing the case for repeaters, we examine what stands between that promise and deployable systems: the duplexing and CSI mechanisms that make the effective channel usable, the wideband delay effects that a narrowband gain model hides, and the sensing dimension, where the same amplification that helps communication actively works against detection. The article closes with a research roadmap rather than a performance claim.}

\textcolor{black}{The remainder of the article is organized as follows. Sec.~\ref{sec: repeaters_introduction} treats the communication aspects of swarm \gls{ra}-\gls{mimo}; Sec.~\ref{sec:isac-repeaters} turns to \gls{isac} and the resulting communication-sensing tradeoffs. Sec.~\ref{sec:outlook} presents the outlook.}

\section{Communication Aspects of Swarm \gls{ra}-\gls{mimo}}\label{sec: repeaters_introduction}
This section separates the practical foundations of \gls{ra}-\gls{mimo} from recent advances in interacting repeater swarms, activation and gain control, duplexing, and wideband frequency diversity. We first summarize the standardized and hardware foundations and then focus on the new results and remaining communication-design challenges. The propagation model introduced in~\Cref{sec: prop_ch} is common to both the communication aspects considered here and the \gls{isac} aspects in~\Cref{sec:isac-repeaters}.

\subsection{Practical Foundations and Recent Advances}
\label{subsec:ncr-standardization}

The standardization of \glspl{ncr} in 5G-Advanced provides a practical baseline for \gls{ra}-\gls{mimo} deployment~\cite{3gpp_ts38106}. The main distinction is one of perspective: cellular standards specify repeater nodes, signals, and procedures, whereas \gls{ra}-\gls{mimo} treats the repeater-assisted paths as part of an effective channel that can be shaped to improve coverage, macro-diversity, and spatial multiplexing. This abstraction preserves centralized multi-antenna processing at the \gls{bs}, but its validity depends on the physical repeater response.

For example, the two directions of a dual-antenna repeater in~\Cref{fig:ra_mimo_config} use different RF chains, including \glspl{lna}, \glspl{pa}, filters, and control circuits. Although the over-the-air propagation channels are reciprocal, mismatched RF responses can make the effective uplink and downlink channels differ in amplitude and phase. Reciprocity calibration is therefore needed for transparent \gls{tdd} operation and reciprocity-based beamforming~\cite{Erik_WCL}. More generally, hardware-aware system models should account for calibration uncertainty, finite dynamic range, \gls{pa} nonlinearities, amplified receiver noise, self-interference leakage, and frequency-dependent filtering. These effects cannot directly be appended after ideal gain optimization because they constrain the useful gain and the repeater output power. Filtering and signal processing also introduce group and processing delays; their wideband consequences are treated in Section~\ref{subsec:wideband-ofdm}. The remaining implementation question is which impairments dominate at the system level and therefore must be estimated, calibrated, controlled, or standardized.

\subsection{Swarm Deployment, Stability, and Practical Control}
\label{subsec:repeater-deployment}

Flexible, plug-and-play deployment is attractive, but the value of each repeater remains geometry-dependent. Well-placed repeaters can illuminate coverage holes, strengthen cell-edge links, increase macro-diversity, and add useful rank to a poorly conditioned MIMO channel. Conversely, a repeater that is weakly connected to the \gls{bs}, far from the intended users, or exposed to strong interference may contribute mainly amplified noise and unwanted signals. Placement and density planning should therefore account for blockage, line-of-sight probability, angular diversity, traffic distribution, and the quality of both hops of each repeater-assisted path.

At swarm scale, density also changes the network dynamics because repeaters hear and amplify one another. Recent work has identified positive feedback as a concern, derived interaction-stability conditions, and analyzed performance within the stable operating region~\cite{Jianan_TWC}. The stability margin depends on the inter-repeater channels and the selected gains, so placement, density, activation, and power control cannot be optimized independently. Activation control can additionally trade fairness and energy efficiency against the number of active artificial-scattering paths~\cite{Emil_repeater}. A practical design should jointly account for the desired signal, amplified noise, interference, output-power limits, and a sufficient stability margin.

This coupling makes the control timescale important. Fast channel-adaptive gain or phase updates may exploit instantaneous \gls{csi}, but require accurate measurements, low-latency signaling, and hardware that stays calibrated and linear over the update interval. Long-term activation maps, sleep modes, and coarse gain control per repeater cluster reduce signaling and implementation complexity, but may sacrifice \gls{se}, fairness, or coverage. Robust policies must determine how much performance is retained under incomplete \gls{csi}, finite control resolution, and delayed updates. Dense deployments can also amplify inter-cell interference or signals belonging to another operator. Band selectivity limits inter-operator coupling, but multi-cell and multi-operator operation may still require coordinated activation, gain limits, or exclusion policies. The resulting research problem is not merely where to install more repeaters, but how to jointly plan and control a stable swarm under realistic information and coordination constraints.

\subsection{Duplexing-Aware Swarm Control and Cross-Link Interference}
\label{subsec:duplexing-csi}

Early \gls{ra}-\gls{mimo} studies mostly assume calibrated \gls{tdd}, where uplink estimated channels are used for downlink precoding via channel reciprocity. This obviates downlink \gls{csi} overhead, but contingent on the repeater-assisted channel being reciprocal after calibration. 
In contrast, \gls{fdd} avoids reciprocity calibration through explicit feedback, at the cost of potentially large overhead. Its advantage is that uplink and downlink repeater gains can be optimized independently, which can improve \gls{se} under certain power and overhead regimes~\cite{Kohei_TDDvsFDD}. This motivates duplexing-aware \gls{ra}-\gls{mimo} studies that quantify when calibrated \gls{tdd}, feedback-based \gls{fdd}, or independent gain control is preferable, accounting for imperfect calibration and finite feedback.

In parallel, dynamic \gls{tdd} and in-band full-duplex operation are also natural extensions for improving time-frequency reuse~\cite{Martin_ICASSP, Mohammadali_ICASSP}. Their key limitation is cross-link interference, which repeaters may further amplify together with desired signals. Therefore, repeater-assisted dynamic duplexing requires joint scheduling, precoding/combining, repeater gain control, and interference-aware \gls{csi} acquisition. The central question is whether repeaters can be controlled so that their spatial and macro-diversity gains outweigh the additional interference they introduce.

{Moreover, in \gls{tdd}-based systems, uplink/downlink switching at the \gls{bs} and users is coordinated while accounting for timing advance. The question then arises as to when the repeater should switch between uplink and downlink operation. If the repeater switches in synchronization with the \gls{bs}, the timing advance applied at the users may result in a timing mismatch between the repeater and the users. To accommodate this mismatch, a longer guard interval may be required, resulting in a reduction in spectral efficiency. This issue must therefore be carefully analyzed in the duplexing design of \gls{ra}-\gls{mimo}.}

\subsection{Wideband, \Gls{ofdm}, Delay, and Frequency-Diversity Effects}
\label{subsec:wideband-ofdm}

Repeaters are often motivated by their ability to add amplified propagation paths and improve the received \gls{snr}, but a narrowband \gls{snr} model is not sufficient for practical wideband systems.
In addition to the propagation delay of the artificial path, RF filters, baseband filters, and signal processing introduce group and processing delays. The resulting effective channel seen by \gls{ofdm}-based systems is modified in both delay and frequency, affecting frequency-domain channel correlation, amplified-noise statistics, achievable rates, and error probabilities. Frequency-dependent repeater gains and colored amplified noise should therefore be modeled with the waveform numerology and receiver equalization.

For conventional \gls{ofdm} reception, the composite channel response should fit within the cyclic prefix. Recent analysis exposes a complementary opportunity within this constraint. When a sufficiently strong repeater-assisted path arrives at a distinct delay while the total channel remains within the cyclic prefix, it can reduce frequency correlation and increase the frequency diversity available to appropriately coded signaling~\cite{Rep_FDiv}. Thus, a delay that would be immaterial in a narrowband model can become a useful channel-shaping resource in a wideband system. The benefit is not automatic: it depends on the path power, delay separation, underlying power-delay profile, signaling, and receiver.

The same mechanism becomes harmful when the excess propagation and processing delay approaches or exceeds the cyclic prefix. Energy outside the cyclic-prefix-supported channel interval causes inter-symbol and inter-carrier interference rather than usable circular convolution, and a high-gain delayed path can then produce severe degradation. Wideband \gls{ra}-\gls{mimo} design must therefore optimize placement, gain, delay, waveform numerology, and receiver processing jointly, rather than maximizing narrowband received power alone. The key open question is when delayed artificial-scatterer paths provide exploitable frequency diversity and when their interference and receiver complexity outweigh that benefit.

\section{Swarm Repeaters for \gls{isac}}\label{sec:isac-repeaters}

\begin{figure}[t]
    \centering    \includegraphics[width=.5\textwidth]{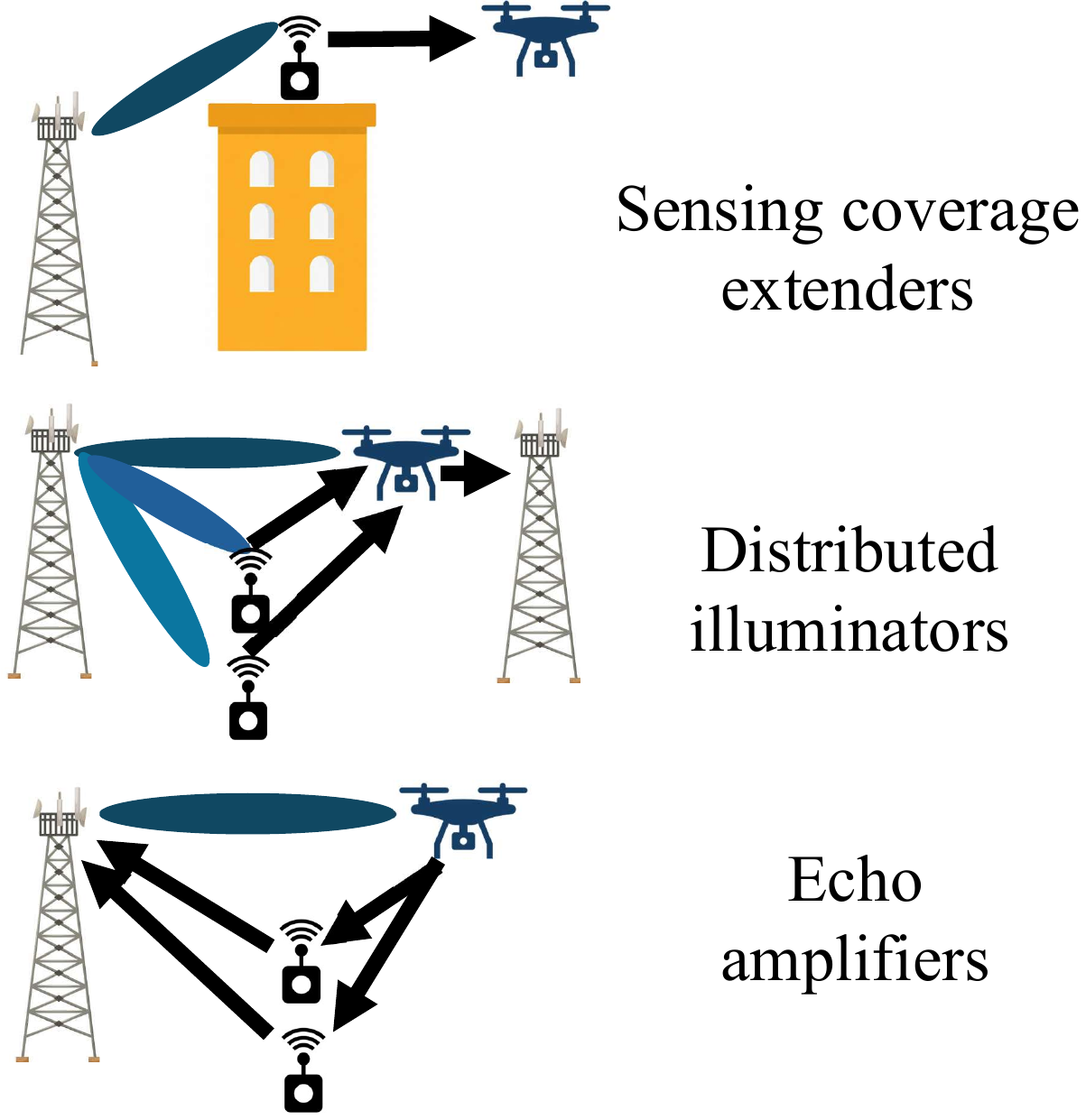}
    \caption{Potential roles of repeaters in \gls{isac}.}
    \label{fig:ra_mimo_ISAC}
\end{figure}

Sec.~\ref{sec: repeaters_introduction} discussed the main considerations and practical insights for densifying networks with swarm repeaters. However, \gls{6g} is expected to evolve beyond connectivity toward \gls{isac}, enabling location-based services and applications involving autonomous systems that heavily rely on environmental awareness, while also leveraging perception to improve the efficiency of the communication system itself. In this context, the role of repeater swarms remains largely unexplored. By acting as band-selective active scatterers with controllable propagation paths, they have the potential to extend sensing coverage to hot-spots, coverage holes, and \gls{los}-blocked regions that \gls{ap}-centric architectures may not easily reach. This can improve target observability and spatial resolution at low infrastructure cost. Yet, repeater settings optimized for communication may not be optimal for sensing. This section examines the role of repeaters in \gls{isac}, the resulting communication-sensing trade-offs, and key design challenges for swarm repeater-assisted \gls{isac}. See~\Cref{fig:ra_mimo_ISAC} for an illustrative example.

\subsection{The Role of Repeaters in \gls{isac}}

In \gls{isac}, repeaters can be exploited as controllable active propagation nodes that reshape both communication and sensing channels. We next discuss the possible roles repeaters can play in enhancing the performance of \gls{isac} systems.

\subsubsection{Repeaters for Coverage Extension, ``\textit{See behind the corners}''}

While \gls{nlos} can be exploited for communications to attain high capacity and spatial diversity by compensating at the receiver side through \gls{mimo}, sensing in \gls{nlos} remains a significant challenge. For radar-like sensing, the scattered power from a target is typically much lower than that of a wave reflected by a large obstacle, leading to higher path-loss for sensing compared to communication. However, sensing can benefit from double reflections collected via passive reflectors with known positions and electromagnetic properties, enabling the gathering of echoes from targets within a desired region of interest~\cite{art:bellini2025}. Translating this principle to repeaters raises challenges due to noise amplification, hardware-induced phases, and radiation in all directions, which makes it harder to associate with a specific geometry. However, under careful calibration and signal processing methods that account for the non-directional radiation and the amplifications, repeaters can potentially overcome passive reflectors.  

\subsubsection{Repeaters as Distributed Illuminators, ``Increasing Target Observability"}

Their controllable gain provides an additional system-level degree of freedom. With proper gain control, repeater swarms can form a distributed sensing support layer that improves target observability and blockage robustness. This requires characterization of AP-repeater, repeater-user, repeater-target, and inter-repeater channels. Since transparent low-complexity repeaters may lack pilot transmission or local channel estimation, channel charting, blind estimation, or network-side inference is needed.

\subsubsection{Repeaters as Echo Amplifiers, ``Enhancing Weak Target Detection"}

A swarm of repeaters can also be deployed to amplify echoes from weak targets in a region of interest, either in monostatic~\cite{art:Jopanya2025} or bistatic settings~\cite{AC_EGL_MIMO_ISAC}. The repeaters' gains, as well as their activation and deactivation, can be optimized to maximize sensing \gls{sinr}, accounting for positive-feedback effects that may degrade performance. Optimal solutions tend to assign a dominant gain to the repeater closest to the target~\cite{art:Jopanya2025}.

\subsection{Communications-Sensing Tradeoff in \gls{ra}-\gls{mimo}}

In \gls{ra}-\gls{mimo} \gls{isac},  several fundamental tradeoffs arise, which must be jointly managed rather than optimized in an independent manner.

\subsubsection{Multipath and propagation artifacts}

Multipath introduced by repeaters acting as scatterers can be exploited for communications to provide diversity and coverage. However, this may introduce adverse effects on sensing, which complicates target discrimination. Particularly, frequency-dependent phase shifts and internal repeater delays can be absorbed as effective multipath in communication channels, but in sensing, these manifest as ghost targets or biased range estimates and degraded Doppler estimation. Nonetheless, stable and known delays can be calibrated to exploit repeaters for enhanced sensing as described in the previous section. However, unknown or time-varying delays must be jointly estimated alongside the sensing scene, which increases the system complexity.

\subsubsection{Dual-function interference} 

In \gls{ra}-\gls{mimo} \gls{isac}, repeaters indiscriminately amplify everything incident on them, including data signals, target echoes, interference, clutter, and noise. This creates a tension when the repeater amplifies and forwards signals to improve link budget and target illumination, but at the same time leaks sensing echoes into communication receivers and communication signals into sensing receivers, thus generating mutual dual-function interference.

\subsubsection{Gain- and density-dependent trade-offs} 

The severity of these effects is governed by repeater gains, density of deployment, AP precoders, geometry, and available \gls{csi}. Low repeater gains limit interference, noise amplification, and feedback instability, but at the cost of reduced coverage and diminished sensing-aperture gains. On the other hand, high gains and dense deployments improve received power and sensing resolution, but risk amplifying clutter and repeater noise, degrading communication SINR.

These tradeoffs require system optimization to be treated jointly across communication rate, sensing accuracy, detection probability, interference leakage, repeater power, and system stability, while accounting for transparent repeaters with limited processing and restricted pilot/\gls{csi} acquisition.
\par
\begin{figure}[t]
    \centering
    \includegraphics[width=0.9\linewidth]{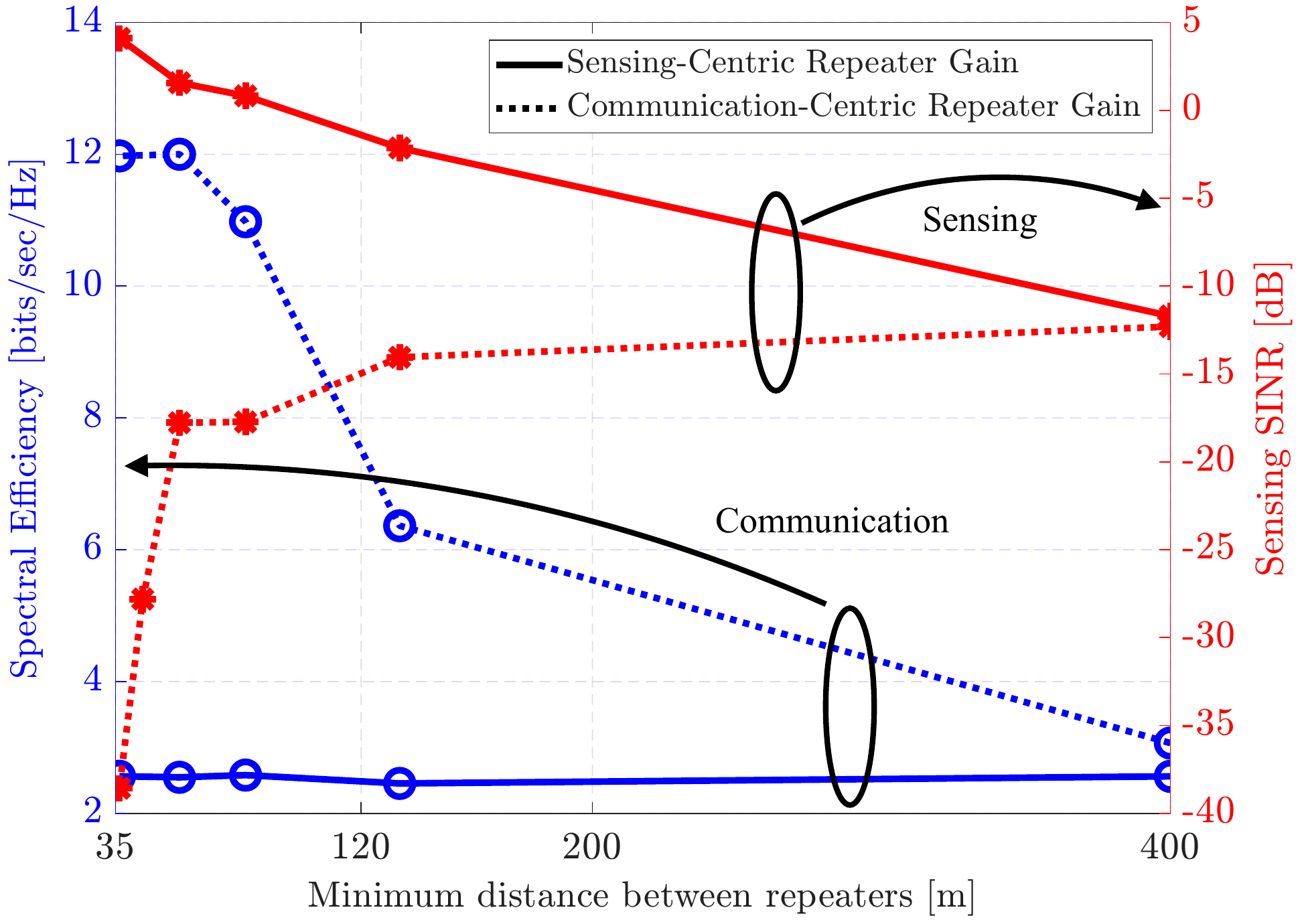}
    \caption{Spectral efficiency of each user and sensing SINR versus minimum inter-repeater distance. Here, we evaluate two different repeater gain strategies: communication-centric and sensing-centric gain.}
    \label{fig:numerical_result}
\end{figure}

{\color{black}To illustrate these tradeoffs, Fig.~\ref{fig:numerical_result} shows the downlink communication \ac{se} of each user
and the sensing \ac{sinr} as functions of the minimum inter-repeater distance, obtained by varying the number of repeaters from $4$ to $144$ within a fixed $400 \times 400$ m cell.
The \ac{bs} is equipped with 128 antennas with 33 dBm transmit power at $15$ GHz, and the target (drone) is located $300$ m from the \ac{bs}, following a setup similar to \cite{art:Jopanya2025}. 
For communication, 8 users are randomly and uniformly distributed within the cell, independent of the repeater density.
The SE is evaluated under zero-forcing precoding, while the sensing SINR follows the same monostatic drone-detection model as in \cite{art:Jopanya2025}.
In the communication-centric system, all repeaters utilize the maximum allowable gain within the stability constraint \cite{Jianan_TWC}. 
On the other hand, the sensing-centric system selects the single repeater that maximizes the sensing SINR, with its gain set to the maximum amplification level \cite{art:Jopanya2025}.
\par
Under communication-centric gain, the SE improves as the deployment becomes denser, but the improvement diminishes beyond a certain density, indicating that simply adding more repeaters yields increasingly marginal gains. On the contrary, the sensing SINR degrades sharply due to accumulated interference and repeater noise. 
Under the sensing-centric gain, this trend is reversed.
The sensing SINR stays high, especially at high density, since a denser deployment offers more candidate repeaters to select from, while the SE remains low and nearly flat, as only a single repeater ever contributes to the communication link.
These results lead to two observations. First, communication benefits from aggregate amplified paths across many repeaters, while sensing benefits from a single well-placed, high-gain path, so a gain policy favoring one objective inherently penalizes the other. Second, communication SE gains diminish beyond a certain density, so simply increasing repeater density mainly consumes power and hardware without commensurate benefit.
Therefore, these observations motivate the future research direction on deployment- and gain-optimization strategies that jointly account for both objectives, as listed further in Sec. \ref{sec:outlook}.}

\subsection{System \& System-level Design}

Beyond the communication-oriented hardware aspects previously discussed, repeater-assisted \gls{isac} introduces additional signal- and system-level challenges because the same repeater simultaneously affects communication and sensing signals~\cite{AC_EGL_MIMO_ISAC}. Active RF chains amplify not only useful signals but also thermal noise, clutter, interference, and hardware distortion, thereby degrading both communication SINR and sensing detection performance. While repeater-originated noise can be partially mitigated through spatial processing when its covariance is known, hardware impairments such as phase noise, PA nonlinearities, I/Q imbalance, and reciprocity-calibration errors reduce coherent integration, distort transmitted waveforms, bias range-angle estimation, and degrade sensing resolution. These effects therefore need to be explicitly incorporated into repeater gain control, waveform design, receiver processing, and calibration procedures rather than treated independently.

At the system level, these impairments are tightly coupled with resource allocation and network operation. In bistatic and multistatic \gls{isac}, access points may dynamically act as transmitters, receivers, or joint \gls{isac} nodes, while repeaters can support communication, sensing, or both. Consequently, repeater activation, gain, \gls{tdd} configuration, \gls{ap} precoding/combining, synchronization, and channel training should be jointly optimized according to communication and sensing objectives. Moreover, user mobility and target dynamics require continuous tracking of \gls{ap}-repeater-user and \gls{ap}-repeater-target channels, whereas dense deployments increase training overhead and computational complexity. Developing scalable signal processing and resource-allocation frameworks that jointly account for these coupled effects remains an important open research direction.

\section{Outlook \& Road Ahead}
\label{sec:outlook}
This article reviewed the physical-layer foundations of \gls{ra}-\gls{mimo}, covering practical deployment considerations, hardware constraints, repeater control, duplexing strategies, channel-state information acquisition, and wideband effects. It further discussed how repeater swarms can support \gls{isac} by extending sensing coverage, improving target observability, and enhancing weak-echo detection, while introducing new communication-sensing tradeoffs arising from amplified noise, clutter, interference, hardware impairments, and stability constraints.

Although recent studies have demonstrated a considerable potential for repeaters, the technology remains at an early stage of development. Many existing results rely on simplified assumptions regarding repeater hardware, propagation, synchronization, channel knowledge, and network operation, whereas practical deployments require scalable, robust, and standardizable solutions that jointly account for these tightly coupled effects. Consequently, future research must progress beyond isolated algorithmic improvements toward integrated communication, sensing, hardware, and network-level designs.

Table~\ref{tab:outlook} summarizes the key open questions identified throughout this article together with promising research approaches for two complementary themes. The first focuses on communication aspects of RA-MIMO, including standardized repeater control, deployment and stability planning, hardware-aware modeling, gain and activation control, duplexing and \gls{csi} acquisition, and wideband channel shaping. The second addresses sensing aspects of \gls{ra}-\gls{mimo}, covering non-line-of-sight sensing, weak-target echo amplification, repeater-induced multipath, communication--sensing interference management, and system-level \gls{isac} design. Collectively, these directions outline a research roadmap toward practical, scalable, and intelligent repeater-assisted wireless networks for future \gls{6g} systems.

\definecolor{TableHeaderColor}{HTML}{D9EAF7}
\definecolor{FirstColColor}{HTML}{F4F8FC}
\definecolor{CommSectionColor}{HTML}{FFF7D6}
\definecolor{SensingSectionColor}{HTML}{EAF6EA}

\newcolumntype{D}{>{\columncolor{FirstColColor}\raggedright\arraybackslash\bfseries}p{2.8cm}}
\newcolumntype{Y}{>{\raggedright\arraybackslash}X}

\begin{table*}[h]
\centering
\footnotesize
\caption{Research Outlook: Key Directions, Open Questions, and Possible Approaches.}
\label{tab:outlook}
\setlength{\tabcolsep}{5pt}
\renewcommand{\arraystretch}{1.13}
\begin{tabularx}{\linewidth}{DYY}
\toprule
\cellcolor{TableHeaderColor}\textbf{Research Direction} &
\cellcolor{TableHeaderColor}\textbf{Key Open Questions} &
\cellcolor{TableHeaderColor}\textbf{Possible Approaches} \\
\midrule

\rowcolor{CommSectionColor}
\multicolumn{3}{l}{\textit{Communication aspects of \gls{ra}-\gls{mimo}}}
\\
\cmidrule(lr){1-3}

Standards and hardware &
Which repeater parameters require network control? &
Specify gain, timing, emissions, noise, linearity, reciprocity, and calibration requirements. \\[4pt]
\cmidrule(lr){1-3}

Swarm planning and control &
How should a stable swarm be deployed and controlled? &
Jointly optimize placement, density, gain, and activation using propagation, traffic, interference, and stability models. \\[4pt]
\cmidrule(lr){1-3}

Duplexing and \gls{csi} &
Which duplexing mode offers the best net gain? &
Compare \gls{tdd}, \gls{fdd}, dynamic \gls{tdd}, and full duplex under calibration, feedback, and cross-link interference. \\[4pt]
\cmidrule(lr){1-3}

Wideband channel shaping &
When does repeater delay help or harm? &
Co-design delay, gain, numerology, cyclic prefix, and receiver processing using frequency-selective channel and noise models. \\

\midrule
\rowcolor{SensingSectionColor}
\multicolumn{3}{l}{\textit{Sensing aspects of \gls{ra}-\gls{mimo}}} \\
\cmidrule(lr){1-3}

\gls{nlos} sensing and coverage extension &
How can repeater swarms enable sensing in \gls{los}-blocked regions?
&
Joint beamforming and gain/activation optimization to form a geometry-aware virtual aperture that can bypass a blocked \gls{bs}-target path while limiting interference elsewhere.

\\[4pt]
\cmidrule(lr){1-3}

Weak-target echo amplification &
How to deploy and optimize a swarm of repeaters to serve as echo amplifiers?&
Optimize repeater gains for sensing, prioritizing repeaters close to the region of interest. \\[4pt]
\cmidrule(lr){1-3}

Repeater-induced multipath effects in sensing performance &
How to account for repeater-induced artifacts when designing \gls{ra}-assisted \gls{isac} systems?
&
Jointly estimate target parameters, repeater delays, and hardware phase offsets to differentiate actual targets from repeater-induced multipath.
\\[4pt]
\cmidrule(lr){1-3}

Dual-function interference management &
How can \gls{ra}-\gls{mimo} \gls{isac} balance communication and sensing performance?&
Jointly optimize \gls{ap} precoders, receive combining, and repeater gains under communication-rate and sensing-detection constraints. \\[4pt]
\cmidrule(lr){1-3}

\gls{ra}-\gls{mimo} \gls{isac} system design &
What are the key design considerations for RA-MIMO ISAC systems? &
Design sensing-aware frame structures with \glspl{ap} acting as transmitters, receivers, or joint \gls{isac} nodes, while optimizing the duplexing mode.  
\\

\bottomrule
\end{tabularx}
\end{table*}

\vspace{-0.5cm}

\bibliographystyle{IEEEtran.bst}
\typeout{}
\bibliography{IEEEabrv.bib, references.bib}
\begin{IEEEbiographynophoto}{DIANA MOYA OSORIO} (Senior Member, IEEE) {is currently an Associate Professor at  Linköping University, Sweden. Her research interests include signal processing for wireless communications and radar systems.}
\end{IEEEbiographynophoto}
\vspace*{-\baselineskip}
\begin{IEEEbiographynophoto}{KOHEI UEDA} (Graduate Student Member, IEEE) is currently pursuing the Ph.D. degree with The University of Electro-Communications, Tokyo, Japan. His research interests include signal processing and communication theory.
\end{IEEEbiographynophoto}
\vspace*{-\baselineskip}
\begin{IEEEbiographynophoto}{ANUBHAB CHOWDHURY} (Member, IEEE) is a post-doctoral researcher at Linköping University,
Sweden. His research interests include signal processing and optimization for next-generation wireless systems.
\end{IEEEbiographynophoto}
\vspace*{-\baselineskip}
\begin{IEEEbiographynophoto}{HIROKI IIMORI} (Member, IEEE) {is a Senior Researcher at Ericsson Research, specializing in 3GPP RAN1 MIMO technologies, wireless communications, signal processing, machine learning, and optimization.}
\end{IEEEbiographynophoto}
\vspace*{-\baselineskip}
\begin{IEEEbiographynophoto}{YUTO HAMA} (Member, IEEE) {is an Experienced Researcher at Ericsson Research with research interests in 3GPP RAN1 standardization, wireless communications, signal processing, and information theory.}
\end{IEEEbiographynophoto}
\vspace*{-\baselineskip}
\begin{IEEEbiographynophoto}{KOJI ISHIBASHI} (Senior Member, IEEE) {is a Professor at The University of Electro-Communications, Japan. His interests are in wireless communications, statistical signal
processing, optimization theory, and information theory.}
\end{IEEEbiographynophoto}
\vspace*{-\baselineskip}
\begin{IEEEbiographynophoto}{ERIK G. LARSSON} (Fellow, IEEE) {is a Professor at Linköping University,
Sweden. His interests are in wireless communications, statistical signal
processing, decentralized machine learning, and network science.}
\end{IEEEbiographynophoto}

\end{document}